%% file: main.tex
\documentclass{article}
\usepackage[T1]{fontenc} 
\usepackage[utf8]{inputenc} 
\usepackage{ismir,amsmath,cite,url}
\usepackage{graphicx}
\usepackage{color}

\usepackage{xcolor}

\usepackage{lineno}

\usepackage{multirow}
\usepackage{booktabs}

\usepackage{makecell}

\usepackage{float}
\usepackage{stfloats}
\usepackage{amssymb} 

\usepackage{algorithm}
\usepackage{algorithmic}

\usepackage{tabularx}

\title{Streaming Piano Transcription 
Based on Consistent Onset and Offset Decoding
with Sustain Pedal Detection}

\oneauthor
{
Weixing Wei$^1$ \hspace{1cm} 
Jiahao Zhao$^1$ \hspace{1cm} 
Yulun Wu$^2$ \hspace{1cm} 
Kazuyoshi Yoshii$^3$
}
{
$^1$Graduate School of Informatics, Kyoto University, Japan \\
$^2$School of Computer Science and Technology, Fudan University, China\\
$^3$Graduate School of Engineering, Kyoto University, Japan\\
{\tt\small \{wei.weixing.23w, zhao.jiahao.56h\}@st.kyoto-u.ac.jp,
yoshii.kazuyoshi.3r@kyoto-u.ac.jp}
}

\def\authorname{W. Wei, J. Zhao, Y. Wu, and K. Yoshii}

\begin{document}

\maketitle

\input{contents/0.Abstract}

\input{contents/1.Introduction}

\input{contents/2.Related_work}

\input{contents/3.Model}

\input{contents/4.Experiments}

\input{contents/5.Conclusion}

\section{Acknowledgments}

This work was partially supported by JST SPRING No. JPMJSP2110, 
JST FOREST No. JPMJFR2270, and JSPS KAKENHI Nos. 24H00742 and 24H00748.





\bibliography{ISMIRtemplate}

%
%
%
%
%

\end{document}

%% file: contents/0.Abstract.tex
\begin{abstract}
This paper describes a streaming audio-to-MIDI transcription method 
that can sequentially translate a piano recording into a sequence of note-on and note-off events. 
The sequence-to-sequence learning nature of this task may call for using a Transformer model, 
which has been used for offline transcription and 
could be extended for streaming transcription with a causal restriction of the attention mechanism. 
We assume that the decoder of this model suffers from the performance limitation.
Although time-frequency features useful for onset detection are considerably different from those for offset detection, the single decoder is trained to output a mixed sequence of onset and offset events without guarantee of the correspondence between the onset and offset events of the same note. 
To overcome this limitation, we propose a streaming encoder-decoder model that uses a convolutional encoder aggregating local acoustic features, followed by an autoregressive transformer decoder detecting a variable number of onset events and another decoder detecting the offset events of the active pitches with validation of the sustain pedal at each time frame. 
Experiments using the MAESTRO dataset showed that the proposed streaming method performed comparably with or even better than the state-of-the-art offline methods while significantly reducing the computational cost.

\end{abstract}


%% file: contents/1.Introduction.tex
\section{Introduction}
\label{sec:intro}

Automatic music transcription (AMT) 
 is a central topic in the field of 
 music information retrieval (MIR),
 which refers to converting a music recording
 into a symbolic musical score (MusicXML format) 
 or a piano-roll representation (MIDI format)
 \cite{AMT-overview/benetos2018}.
It has remarkably been improved
 with the technical progress of deep learning techniques and 
 the public availability of large-scale music datasets.
In this paper,
 we focus on streaming audio-to-MIDI AMT 
 because it remains relatively unexplored 
 unlike streaming automatic speech recognition (ASR)
 \cite{rnnt-asr:conf/icassp/GravesMH13,mocha/iclr/ChiuR18,real-time-streaming-asr-chen2021developing}
 and forms the basis of real-time music applications
 such as performance evaluation and interactive jam session.
The previous research in \cite{ismir/KwonJN20/Autoregressive} 
applied auto-regressive convolutional recurrent neural network (CRNN) frame-by-frame for piano transcription. 
The auto-regressive CRNN model can be easily adapted for the online scenario \cite{ismir/jeong2020real}.
But the transcription performance for note offsets still has significant room for improvement.

\begin{figure}[t]
\centering
\includegraphics[scale=0.7]{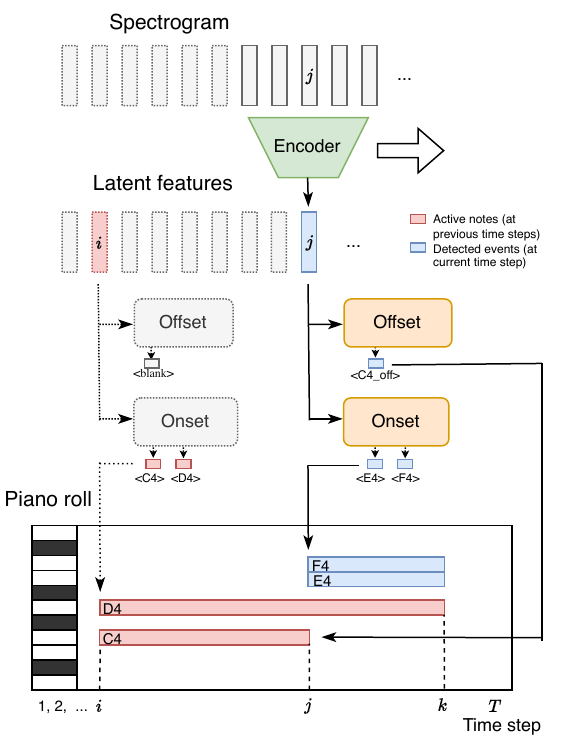}
\vspace{-4mm}
\caption{An overview of the proposed streaming audio-to-MIDI piano transcription method
aware of onset-offset correspondence.}
\vspace{-3mm}
\label{fig:model_overview}
\end{figure}

Inspired by the sequence-to-sequence learning for ASR,
 many studies on AMT have recently attempted 
 to use the Transformer \cite{attention-is-all-you-need/nips/VaswaniSPUJGKP17}
 by serializing the polyphonic information of the estimation target
 \cite{transformer,MT3/iclr/GardnerSMHE22}.
AMT is essentially different with ASR
 in a sense that 
 the onsets, durations, and pitches of musical notes
 should be estimated, 
 while the temporal information of output tokens 
 (e.g., words and characters)
 is not considered in ASR.
For audio-to-MIDI piano transcription,
 one may define the input and output of the Transformer 
 as a sequence of raw audio features
 (e.g., mel and constant-Q spectrograms)
 and a sequence of note-on and note-off events 
 sorted in time and pitch, respectively.
The performance of this naive approach, however, 
 is potentially limited.
Despite the significant differences in features needed for detecting onsets and offsets,
the Transformer decoder estimates these events in a mixed manner.
In addition, the correspondence 
 between the onset and offset events of the same note
 is not guaranteed.

For streaming AMT,
 one can use the \textit{causal} Transformer
 that restricts the self-attentive region 
 to a certain number of past frames,
 which could reduce
 the computational cost of the basic self-attention mechanism
 that increases quadratically with the input length.
Nonetheless, 
 due to the strong coupling between note events,
 Transformer-based transcription methods 
 often underperform the state-of-the-art frame-level methods 
 \cite{hft-transformer-toyama2023, TF-Perceiver/icassp/LuWH23},
 especially in offset detection and velocity estimation. 


To overcome these limitations,
 we propose a streaming audio-to-MIDI piano transcription method
 based on a novel encoder-decoder architecture (Fig.~\ref{fig:model_overview}).
The encoder is implemented with a convolutional neural network (CNN) 
 that sequentially 
 aggregates latent features from local regions of an input piano recording.
The two Transformer decoders that operate framewise
 are then separately used for detecting a variable number of onset events 
 and offset events for the active pitches
 with guarantee of onset-offset correspondence.
For further improvement,
 the offset decoder is trained 
 to judge the activation of the sustain pedal
 in a way of multitask learning.



The main contribution of this study is 
 to develop an efficient streaming encoder-decoder model 
 and pave a way for interactive and responsive applications
 based on real-time music transcription.
We experimentally show that our method performs comparably with a state-of-the-art offline transcription method and outperforms existing sequence-to-sequence transcription methods.





%% file: contents/2.Related_work.tex
\section{Related Work}
\label{sec:related}

This section reviews related work on automatic music transcription 
 and sequence-to-sequence transcription.

\subsection{Automatic Piano Transcription}

Automatic piano transcription (APT) is the most popular form of AMT.
Early methods rely on handcrafted features and rule-based algorithms 
 \cite{Non-Negative, Abdallah, VincentBB10, NamNLS11}, 
 while modern methods use deep learning models
 such as CNNs 
 \cite{on-the-potential/ismir/KelzDKBAW16, 
 deep-adsr/icassp/KelzBW19, SigtiaBD16, Onsets_Frames}, 
 recurrent neural networks (RNNs) 
 \cite{BockS12, modeling-temporal/icml/Boulanger-LewandowskiBV12}, 
 and transformers 
 \cite{exploring-transformer/icassp/OuGBHW22, harmonic-transformer/icassp/wu}.
In APT,
 the framewise transcription has still been the mainstream approach
 due to its superior performance and accuracy
 \cite{hft-transformer-toyama2023, hppnet-Wei2022}. 
In this approach,
 audio features 
 such as short-time Fourier transform (STFT) spectrograms 
 are mapped to a binary matrix of dimensions $T\times N$
 indicating the presence of pitches over time frames,
 where $T$ represents the number of frames 
 and $N$ the number of pitches.
Early transcription methods, mostly based on CNNs, 
 perform comparably at the frame level 
 but underperform in term of note-level.

Onsets and Frames \cite{Onsets_Frames}
 is a major breakthrough in APT
 that learns to sequentially predict 
 note onsets and pitches
 in a multitask framework.
To improve the performance,
 a music language model (MLM) 
 based on a bidirectional long short-term memory (BiLSTM) network
 is used for modeling the temporal dependency of musical notes.
This study has triggered many extensions.
Kong~et~al.~\cite{high_resolution}, for example,
 proposed a high-resolution piano transcription (HPT) model
 that simultaneously deals with onset, offset, velocity, 
 and frame prediction tasks. The predicted velocities are used as conditional information to predict onsets, and the predicted onsets and offsets are used to predict frame-wise pitches, forming a hierarchical structure.

Our previous work \cite{hppnet-Wei2022} proposed
 HPPNet that uses harmonic dilated convolution
 for constant-Q transform (CQT) spectrograms
 and an enhanced frequency grouped LSTM (FG-LSTM) as a MLM. 
This model exhibits improved performance 
 in both frame-level and note-level predictions. 
To capture long-term temporal and spectral dependencies, 
 Toyama et al.~\cite{hft-transformer-toyama2023}
 proposed a two-level hierarchical frequency-time transformer 
 (hFT-Transformer)
 and achieved the state-of-the-art performance on the prediction of note with offset and velocity.

\subsection{Sequence-to-Sequence Transcription}

Sequence-to-sequence models are able to learn a mapping 
 between input and output sequences of variable lengths
 and have actively been investigated
 in many fields 
 such as natural language processing (NLP) 
 and automatic speech recognition (ASR).
Such models have recently been implemented 
 with the Transformer or the self-attention mechanism
 due to its excellent performance.
Awiszus et~al.~\cite{awiszus2019automatic},
 for example, proposed a piano transcription model 
 based on an LSTM and a Transformer 
 for frame-level multi-pitch estimation.
The performance of this method, however, is limited
 due to the lack of training data and using improper relative time shifts.

Inspired by this study, 
 Hawthorne et al.~\cite{transformer} 
 proposed a note-level piano transcription model
 that uses Transformer encoder and decoder
 in a way similar to the T5 model~\cite{T5-RaffelSRLNMZLL20}.
The encoder extracts latent features 
 from an input spectrogram
 and the decoder refers to the input 
 in an autoregressive manner, 
 and the token with the highest probability 
 is selected at each frame.
This method achieved promising performance 
 on the MAESTRO dataset
 and was later extended to multi-track music transcription\cite{MT3/iclr/GardnerSMHE22}. 
However,
 this sequence-to-sequence transcription method still faces limitations. 
It encodes all types of note events and 
 absolute time location of each event
 into a single sequence.
This increases the complexity of sequence-to-sequence transformation and
 also constrains the length of the input sequence.


%% file: contents/3.Model.tex
\section{Proposed Method}
\label{sec:model}


This section explains 
 the proposed method of streaming audio-to-MIDI 
 piano transcription
 based on a single encoder and onset and offset decoders.

\subsection{Streaming Transcription}

As shown in Algorithm \ref{alg:streaming_transcription},
 the model takes a spectrogram
 $\textbf{X} \in \mathbb{R}^{T \times F_{i}}$ as input,
 where $T$ represents the number of frames
 and $F_i$ represents the number of frequency bins.
It outputs an onset sequence list $\mathbf{Y}$ 
 and an offset sequence list $\mathbf{\overline{Y}}$, 
 where each element $\textbf{Y}_t$ in $\mathbf{Y}$ 
 represents the detected onsets sequence of frame $t$ 
 with sequence length $k_t$, 
 and each element $\overline{\textbf{Y}}_t$ in $\mathbf{\overline{Y}}$ 
 represents the detected offsets sequence 
 with sequence length $n_t$ in frame $t$.

\begin{figure}[t]
\centering
\includegraphics[scale=0.63]{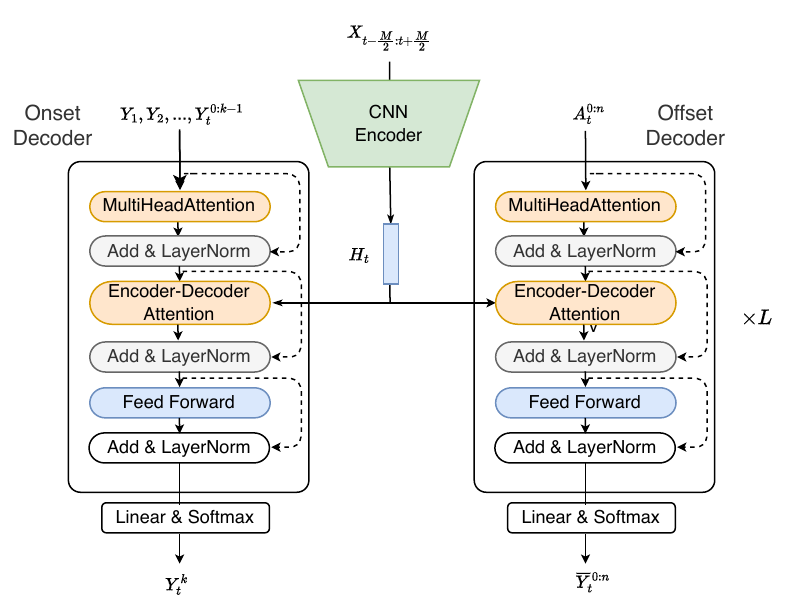}
\vspace{-6mm}
\caption{The implementation of the streaming transcription model
that uses one encoder for latent feature extraction 
and two decoders for onset and offset detection.}
\label{fig:model_architecture}
\vspace{1mm}
\end{figure}

The model consists of one encoder and two decoders
 (Fig.~\ref{fig:model_architecture}).
The encoder is implemented with a CNN 
 that efficiently extracts and aggregates local features 
 from the audio spectrogram $\textbf{X}$. 
The two separate decoders are then used at each frame
 for detecting a variable number of onset times
 and judging the offset of the detected notes
 by focusing on different aspects of the latent features.

More specifically, at each frame $t$, 
 the encoder takes as input 
 the audio spectrogram around frame $t$
 with a receptive field of a fixed size $M$ 
 and outputs a hidden embedding sequence 
 $\mathbf{H}_t\in \mathbb{R}^{F_h \times D}$ 
 in the frequency domain with a sequence length of $F_h$ 
 and the hidden embedding size of $D$. 
In addition,
 positional encodings are incorporated 
 into the encoder hidden states $\mathbf{H}_t$. Then the decoders receive $\mathbf{H}_t$ with the cross attention (encoder-decoder attention).

\input{algorithm/streaming_transcription_encoder_decoder}
\input{algorithm/test}

For onset detection, 
 the onset sequence $\mathbf{Y}_t$ at frame $t$ 
 is initialized with the beginning-of-sequence token (BOS). 
The onset events are then detected
 using the onset decoder $\text{Decoder}_{on}$ iteratively 
 until the end-of-sequence token (EOS) is obtained, 
 considering the current encoder hidden state $\mathbf{H}_t$, 
 the onset sequences $\mathbf{Y}_{1:t-1}$ detected in previous times, 
 the current onset sequence at frame $t$, 
 and decoder positional encodings. 
The detected onset events are finally added 
 to the active onsets set $\mathbf{A}$.
The process is repeated throughout the input sequence $\mathbf{X}$.

The offset events are detected using the offset decoder $\text{Decoder}_{off}$, 
 considering the current encoder hidden state $\mathbf{H}_t$, 
 the active onsets set $\mathbf{A}$, and decoder positional encodings.
Then active onsets corresponding to the detected offsets 
 are removed from $\mathbf{A}$ indicating the end of notes.
It should be emphasized that the offset decoder does not perform sequence prediction. 
Instead, it predicts the offset for each onset
 that has been activated in the past time steps all at once.


\subsection{Encoder}


The encoder is based on the harmonic dilated convolution 
 originally used for HPPNet \cite{hppnet-Wei2022}
 and uses the same configuration proposed
 for the acoustic model of HPPNet. 
It extracts local acoustic features 
 with a fixed receptive field and feeds them to the decoders. 
There are three sets of convolutional layers 
 with different kernel sizes: 
 three layers with a kernel size of $7\times7$, 
 one harmonic dilated convolution layer 
 with a kernel size of $1\times3$, 
 and five layers with a kernel size of $5\times3$. 
The resulting receptive field in the time dimension is $M = 39$.

For streaming piano transcription, 
 we use the shifting window approach 
 for sequentially feeding an input spectrogram to the encoder.  
Instead of feeding the entire spectrogram at once, 
 we segment it into smaller chunks or windows 
 to simulate real-time processing. 
These windows are shifted along the time axis, 
 allowing the model to gradually analyze incoming audio data.
We define the size of each window 
 based on the desired temporal context for transcription. 
Typically, smaller window sizes facilitate faster processing 
 but may sacrifice some contextual information, 
 whereas larger window sizes provide more context 
 but may introduce latency.
To ensure continuity of transcription 
 and avoid information loss at window boundaries, 
 we apply overlap between consecutive windows. 

\subsection{Decoder}

Both the onset and offset decoders are
 the same as the decoder of T5 \cite{T5-RaffelSRLNMZLL20} 
 (Fig.~\ref{fig:model_architecture}).
In the decoder architecture, 
 the embedding size is set to $D_{dec} = 256$, 
 and decoder layers to $L = 6$,
 attention head number to $N_{head} = 8$.
The multi-layer perceptron (MLP) dimension 
 is set to $D_{mlp} = 1024$. 
 A maximum decoder sequence length $N_{seq} = 64$. 
The length of the decoder output varies
 with the number of activated onsets.
During the training phase,
 we use padding and masking 
 to fix the output tokens length of offset decoder to 16.
 

\subsection{Consistent Decoding}

Existing piano transcription models 
 that applied onset and offset detection \cite{high_resolution, MT3/iclr/GardnerSMHE22} 
 often face issues with mismatched detected onsets and offsets. 
This is due to the little constrains 
 in the detection processes for onsets and offsets.
Although this issue can be addressed with post-processing methods,
 we prefer to solve it end-to-end within the model. 
Our proposed architecture makes a constriction to the offset decoder
 to detect offsets for detected onsets only, 
 and also detects sustain pedal release events
 to improve performance of note offsets detection.

The onset decoder sequentially outputs onset events
 in an autoregressive manner
 while the offset decoder detects 
 all the offset events at once 
 for the active notes detected by the onset decoder
 with judgement of the sustain pedal. 
If the offset event for an active note
 is not detected at the current frame, 
 a special token $BLANK$ is obtained
 as described in Section~\ref{sec:outputs}. 
The onset decoder considers 
 only notes detected
 in the past and current frames.
The sustain pedal plays a crucial role 
 in expressive piano performance
 and considerably affects offset detection.
The lifting time of the sustain pedal 
 is highly relevant to the absolute offset times
 and thus determines the duration and decay characteristics
 of musical notes. 
 
The input of the onset decoder in each step at frame $t$
 consists of the onset tokens 
 detected in the previous step 
 and the onset tokens detected at previous frames. 
This enables to capture 
 long-term dependency between musical notes. 
By incorporating information from previous frames, 
 the decoder can better understand 
 the context of the current onset detection
 and facilitate 
 the recognition of typical patterns and structures 
 in the music sequence over time. 





%% file: algorithm/streaming_transcription_encoder_decoder.tex
\begin{algorithm}[t]
\caption{Streaming piano transcription. The length of output onset sequence equals to the number of the detected onsets, while the length of offset sequence has an additional output for pedal offset indexed as 0.}
\begin{algorithmic}[1]
\STATE \textbf{Input}: Source sequence $\mathbf{X} = (x_1, x_2, ..., x_T)$
\STATE \textbf{Output}: 
\STATE Onset sequence $\mathbf{Y} = (Y_{1}^{1:k_1}, Y_{2}^{1:k_2}, ..., Y_{T}^{1:k_T})$
\STATE Offset sequence $\mathbf{\overline{Y}} = (\overline{Y}_{1}^{0:n_1}, \overline{Y}_{2}^{0:n_2}, ..., \overline{Y}_{T}^{0:n_T})$

\STATE \textbf{Parameters}: 
\STATE Receptive field of encoder: M
\STATE Initialize positional encodings: $\mathbf{PE}_{\text{enc}}$ and $\mathbf{PE}_{\text{dec}}$
\STATE Initialize active onsets set: $\mathbf{A} = \{\}$
\FOR{$t = 1$ \textbf{to} $T$}

    \STATE $H_t \leftarrow \text{Encoder}(X_{t-\frac{M}{2}:t+\frac{M}{2}})$
    \STATE $H_t \leftarrow H_t + \mathbf{PE}_{\text{enc}}$
    \STATE \text{// Offset decoder}
    \STATE $n_{t} \leftarrow \mathbf{A}.\text{size}()$
    \STATE $\overline{Y}_{t} \leftarrow \text{Decoder}_{off}(H_t, \mathbf{A},\mathbf{PE}_{\text{dec}})$
    \STATE Delete onsets in $\mathbf{A}$ corresponding to offsets in $\overline{Y}_{t}$

    \STATE \text{// Onset decoder}
    \STATE $k_t \leftarrow 0 $
    \STATE $Y_t^{k_t} \leftarrow \text{BOS}$
    \STATE $y \leftarrow \text{BOS}$
    \WHILE{\text{$y$ != EOS}}
        \STATE $y \leftarrow \text{Decoder}_{on}(H_t, Y_{1:t-1}, Y_t^{0:k_t},\mathbf{PE}_{\text{dec}})$
        \IF{y == \text{EOS}}
            \STATE \textbf{break}
        \ENDIF
        \STATE $k_t \leftarrow k_t + 1 $
        \STATE $Y_t^{k_t} \leftarrow y$
        
    \ENDWHILE
    \STATE $\mathbf{A}.\text{add}(Y_t)$
    
\ENDFOR

\end{algorithmic}
\label{alg:streaming_transcription}
\vspace{-1mm}

\end{algorithm}

%% file: algorithm/test.tex

%% file: contents/4.Experiments.tex
\section{Evaluation}
\label{sec:experiments}

This section reports a comparative experiment 
 conducted for evaluating the performance 
 of the proposed and conventional piano transcription methods.





\input{tables/output_targets}

\input{tables/maestro_test_results}
\input{tables/Seq2Seq}
\input{tables/ablation_study}

\subsection{Experimental Conditions}

We explain the dataset used for evaluation
 and the input and output data of the proposed method.

\subsubsection{Dataset}

We used the MAESTRO dataset V3.0.0~\cite{maestro} 
 composed of about 200 hours of virtuosic piano performances captured with fine alignment 
 between note events and audio recordings.
The split of the dataset
 into training, validation, and test sets
 was defined officially.
The validation set was used 
 for selecting the best-performing trained model 
 based on its performance on unseen data.
The dataset also provides information
 about the states 
 (on or off) of the sustain pedal. 
The pedal information is crucial 
 for accurately transcribing piano performances
 as it affects the \textit{actual} 
 durations and offset times of sustained notes.

\subsubsection{Input}


The original audio recordings
 were resampled with a sampling rate of 16 kHz.
To increase the variation of the training data 
 and reduce the memory footprint, 
 10-sec segments were randomly
 clipped from the recordings
 and the CQT spectrograms were computed on the fly
 with the nnAudio library~\cite{nnAudio}. 
We used the CQT for its capability
 of capturing both higher and lower-frequency components 
 in the logarithmic frequency domain
 suitable for analyzing music signals.
The lowest frequency was set to 27.5 Hz
 corresponding to the lowest key
 of the standard 88-key piano. 
One octave was divided into 48 frequency bins 
 and the total number of frequency bins was 352.
This ensures a fine frequency resolution 
 over the entire audible frequency range. 
The hop length was set to 320 samples (20 ms), 
 taking the balance between 
 the time resolution and the computational efficiency.
After obtaining the CQT spectrogram, 
 the amplitude values were converted to decibels (dB) 
 using transforms available in the torchaudio library. 

\subsubsection{Output}
\label{sec:outputs}



The vocabulary of output tokens used in our study
 was the same as that used for the music transformer 3 (MT3) \cite{transformer,MT3/iclr/GardnerSMHE22}
 except that time location tokens were not used.
This contributes to reducing the length of the output sequence
 and stabilizing the training.
The output vocabulary consists of the following tokens:
\begin{description}
  \setlength{\parskip}{1pt} 
  \setlength{\itemsep}{1pt} 
    \item[Onsets and offsets (128+128 tokens)]
    Each token represents the presence of an onset or offset of the corresponding pitch given as a MIDI note number. 
    \item[Pedal states (2 tokens)] 
    Two tokens representing
    the presence and absence of the sustain pedal.
    \item[BLANK (1 token)] A special token representing silence or absence of any musical event.
    \item[BOS and EOS (2 tokens)] Special tokens representing the beginning and end of the output sequence.
\end{description}

The onset decoder and offset decoder
 both need only part of the vocabulary. 
But we kept the full vocabulary for all decoders
 to maintain consistency in the model architecture,
 regardless of whether there is only one decoder or multiple decoders.
We set the length of each onset and offset events into 2 frames. 
During the transcription process,
 if consecutive onsets or offsets were detected,
 we only kept the first one
 and discard the duplicates. 
To estimate note events 
 from the output of the decoders 
 we used a simple greedy regression algorithm.
We then selected
  the nearest corresponding offsets after the onsets to determine the duration of the notes.
If offsets were not detected, 
 we selected the nearest pedal offset as the offsets for the notes or a maximum duration of 4 seconds.

\subsubsection{Training}



We used the cross entropy loss for training the proposed model.
It represents 
 the negative log-posterior probability over output tokens
 for the ground-truth annotation.
For optimization, 
 we utilized the AdamW optimizer
 \cite{adamw-iclr-LoshchilovH19}, 
 which is a variant of the Adam optimizer 
 with weight decay regularization. 
The mini-batch size was set to 16
 and the learning rate was set to 6e-4. 
A dropout rate of 0.1 was applied 
 to the decoder layers to prevent overfitting. 
Training was iterated for 200,000 steps with early stopping.


\subsubsection{Metrics}




The performance of piano transcription
 was evaluated 
 with the mir\_eval library\cite{mir_eval}
 in terms of
 the precision and recall rates and F1 score
 at the frame and note levels.
In the note-level evaluation,
 an estimated note was judged as correct
 if its onset time was detected correctly
 or if both the onset time and duration were estimated correctly.
The error tolerance in onset estimation
 was set to 50 msec as in many studies.
The error tolerance in duration estimation
 was set to the larger of 50 msec 
 or 20\% of the ground-truth duration.
These metrics were averaged
 over the test set.
 

\subsection{Experimental Results}

We report the experimental results
 obtained through comparative and ablation studies.

\subsubsection{Comparison with Existing Methods}

We conducted a comprehensive experiment that compared
 our method with state-of-the-art methods
 such as frame-level and event-level transcription methods
 (Table \ref{tab:maestro}).
We found that our method achieved competitive performance
 and surpassed an event-level transcription method
 named Semi-CRFs in terms of both the note-level F1-scores
 with and without duration evaluation.
This superiority indicates 
 the robustness of our method 
 in capturing the musical onset events 
 and their corresponding offsets. 



\subsubsection{Sequence to Sequence Transcription}

For comparison, 
 we tested the generic transformer-based 
 sequence-to-sequence transcription model~\cite{transformer}
 (Table \ref{tab:seq2seq}).
Audio recordings were split into segments of 4088 msec 
 to be fed to the encoder. 
Since different segments were transcribed independently, 
 the long-term correlation between note events
 is hard to learn from the data.
Moreover, increasing the segment length 
 would exponentially increase 
 the computational complexity of the self-attention layers.
It would increase the number of absolute-time-location tokens
 and further complicates 
 the estimation of time locations for note events.

Thanks to the streaming encoder-decoder architecture, 
 the proposed model kept the actual input length constant
 and significantly reduced the computational complexity
 of the self-attention layers.
The length of the encoder input was set to 39 
 and the maximum length of the decoder output was set to 64. 
This enables the processing of variable-length audio recordings
 without the need for segmentation 
 and offers the potential for real-time transcription.
Compared with the generic model, 
 our streaming model showed better performance 
 in terms of the note-level F1-scores
 with and without duration evaluation.
This indicates the potential application 
 to streaming and sequence-to-sequence 
 music transcription scenarios.

\subsubsection{Latency}

The latency of a streaming model refers to
 the gap between the actual time of an onset or offset event
 and the time of the event output.
Putting the actual computational speed aside,
 the latency of a non-streaming model
 is equal to the length of the input sequence
 because the whole sequence needs to be processed
 for generating outputs. 
In contrast, for streaming models,
 the latency is equal to the length of future frames 
 in the input data stream. 

In Table \ref{tab:seq2seq}, 
 our streaming model
 had a latency of 380 msec.
The CNN-based encoder takes 19 future frames 
 and 19 past frames as input. 
Even with a short input context,
 the streaming model still achieved
 competitive performance on piano transcription.
This indicates that onset and offset events
 could be detected without heavily relying 
 on long-term dependency of acoustic features.


\subsubsection{Ablation Study}

To verify the effectiveness of sustain pedal detection
 and that of the separated decoders 
 for onset and offset detection,
 we conducted an ablation study.
Besides the proposed model,
 we trained a model without pedal detection
 and another model that uses a single decoder
 for onset, offset, and pedal detection. 
The training and evaluation
 were performed in the same way. 

Table \ref{tab:ablation_study}
 shows the performances of the compared methods. 
We found that removing the pedal detection
 slightly decreased the note-level F1-score 
 without duration estimation,
 but significantly degraded
 the note-level F1-score with duration estimation. 
This suggests that pedal detection
 plays a crucial role
 in estimating note durations. 
Similarly, using a single decoder
 for both onset and offset detection
 degraded both the note-level F1-scores 
 with and without duration estimation,
 compared with the proposed model.
This demonstrated the effectiveness
 of incorporating pedal detection
 and a separated decoder
 for onset and offset prediction
 for better piano transcription.


%% file: tables/output_targets.tex
\begin{table}[t]
\small
\begin{tabular}{c|c}
\hline
Time & Target Tokens \\
\hline
1         & {\color{red}<EOS>} {\color{blue}<blank>}             \\
2         &  {\color{red}<EOS>} {\color{blue}<blank>}             \\
...       & \\
i         & {\color{red}<C4><D4><EOS>} {\color{blue}<blank>}    \\
i+1       & {\color{red}<C4><D4><EOS>} {\color{blue}<blank><blank><blank>} \\
i+2       & {\color{red}<EOS>} {\color{blue}<blank><blank><blank>} \\
...       & \\
j         & {\color{red}<E4><F4><EOS>} {\color{blue}<blank><C4\_off><blank>} \\
...       & \\
k         & {\color{red}<EOS>} {\color{blue} <pedal\_off><D4\_off><E4\_off><F4\_off>} \\
\hline

\end{tabular}
\caption{Target tokens for onset decoder(red) and offset decoder(blue).}
\label{tab:target_tokens}
\end{table}




% ...




% ...


% ...


% ...


%% file: tables/maestro_test_results.tex
\begin{table*}[t]
 \begin{center}
 \resizebox{.85\width}{!}{
 \renewcommand{\arraystretch}{1.3}
 \begin{tabular}{l c |c c c c c c c c c }
 \toprule[2pt]
 \multirow{2}{*}{Model} & \multirow{2}{*}{Params} & \multicolumn{3}{c}{Frame-level} & \multicolumn{3}{c}{Note-level (onset only)} &
 \multicolumn{3}{c}{Note-level (onset + duration)} \\
 \cmidrule(r){3-5} \cmidrule(r){6-8} \cmidrule(r){9-11} 
 
 & & P (\%) & R (\%) & F1(\%) & P (\%) & R (\%) & F1(\%) 
 & P (\%) & R (\%) & F1(\%) \\
 
\midrule[1.5pt]

Onsets \& Frames \cite{maestro} & 26M & 92.11 & 88.41 & 90.15 & 98.27 & 92.61 & 95.32 &
82.95 & 78.24 & 80.50  \\


Semi-CRFs \cite{skip_frame} & 9M & 93.79 & 88.36 & 90.75 & 98.69 & 93.96 & 96.11 & 
90.79 & 86.46 & 88.42 \\

HPPNet-sp \cite{hppnet-Wei2022} & 1.2M & 92.79 & 93.59 & \underline{93.15} & 98.45 & 95.95 & \underline{97.18} & 84.88 & 82.76 & 83.80  \\

hFT-Transformer \cite{hft-transformer-toyama2023} & 5.5M & 92.82 & 93.66 & \pmb{93.24} & 99.64 & 95.44 & \pmb{97.44} & 92.52 & 88.69 & \pmb{90.53} \\



\hline
Streaming Seq2Seq (ours) & 16M & 91.91 & 91.73 & 91.75 & 98.30 & 94.83 & 96.52 & 91.08 & 87.89 & \underline{89.44}  \\
 
 \bottomrule[2pt]
 \end{tabular}
 }
\end{center}
\vspace{-5mm}
 \caption{The transcription performances of the existing and proposed methods on MAESTRO V3.0.0 test set.} \label{tab:maestro}
 \vspace{2mm}
\end{table*}

%% file: tables/seq2seq.tex
\begin{table*}[htbp]
\begin{center}
\resizebox{0.9\width}{!}{
\begin{tabularx}{1\textwidth}{l|c|>{\centering\arraybackslash}X|>{\centering\arraybackslash}X|c|c|>{\centering\arraybackslash}X}
 \toprule[1.5pt]
Model & Segment & Encoder Input Seq-Length & Decoder Output Seq-Length & Latency & Note F1 & Note w/ Offset F1 \\
  \midrule[1.2pt]
  Seq2Seq \cite{transformer} & 4088 ms & 511 & 1024 & 4088 ms & 96.01 & 83.94 \\
  Streaming Seq2Seq (ours) & - & 39 & 64 & 380 ms & \pmb{96.52} & \pmb{89.44} \\
 \bottomrule[1.5pt]
\end{tabularx}

}
\vspace{-1.5mm}
\caption{The transcription performances of sequence-to-sequence transcription models on MAESTRO V3.0.0 test set.} \label{tab:seq2seq}
 
\end{center}
\end{table*}

%% file: tables/ablation_study.tex
\begin{table*}[t]
 \begin{center}
 \resizebox{.85\width}{!}{
 \renewcommand{\arraystretch}{1.3}
 \begin{tabular}{c c c c |c c c c c c c c c}
 \toprule[2pt]
 \multirow{2}{*}{Decoder} &\multirow{2}{*}{Onset} & \multirow{2}{*}{Offset}& \multirow{2}{*}{Pedal} & \multicolumn{3}{c}{Note-level (onset only)} &
 \multicolumn{3}{c}{Note-level (onset+duration)}   \\
 
 \cmidrule(r){5-7} \cmidrule(r){8-10} \cmidrule(r){11-13} 
 
 & & & &  P (\%) & R (\%) & F1(\%) 
 & P (\%) & R (\%) & F1(\%) \\
 
\midrule[1.5pt]

1 & \checkmark & \checkmark & \checkmark & 98.32 & 93.36 & 95.73 & 89.91 & 85.41 & 87.56  \\

2 & \checkmark & \checkmark &  &  98.23 & 94.75 & 96.44 &
88.11 & 85.00 & 86.51  \\

2 & \checkmark & \checkmark & \checkmark & 98.30 & 94.83 & \pmb{96.52} & 91.08 & 87.89 & \pmb{89.44} \\

 \bottomrule[2pt]
 \end{tabular}
 }
\end{center}
\vspace{-5mm}
 \caption{Ablation study on MAESTRO V3.0.0 test set.  } \label{tab:ablation_study}
\end{table*}


%% file: contents/5.Conclusion.tex
\section{Conclusion}
\label{sec:conclusion}

In this paper, 
 we have presented a novel streaming
 audio-to-MIDI piano transcription method.
We tackled an open problem
 of detecting note onset and offset events
 from a piano recording in an online manner.
Our method is based on
 a streaming encoder-decoder architecture
 that combines a convolutional encoder
 for aggregating local acoustic features
 with separate transformer decoders
 for detecting onset and offset events at each time step
 while validating the use of the sustain pedal.

In extensive experiments with the MAESTRO dataset,
 our method attained competitive performance,
 compared with the state-of-the-art offline methods.
Our model also outperformed
 the generic transformer-based sequence-to-sequence model
 in terms of both accuracy and latency.
The ablation study showed the effectiveness
 of incorporating pedal detection
 and that of using the separated decoders
 for onset and offset detection.
Our method uses a limited number of incoming frames
 for detecting the onset and offset events
 and paved a way for latency-critical practical applications.
We achieved a system latency of 380 msec
 and plan to thoroughly investigated 
 the trade-off between the latency and the transcription performance.
Additionally, decoding every frame may not be necessary. 
Some scenarios might not require such high temporal precision.
The setting of the time step also requires further exploration for real-time scenarios.